\documentclass{ws-ijmpa}

\usepackage[super]{cite}
\usepackage{xcolor}
\usepackage[verbose,hypertexnames=false]{hyperref}
\hypersetup{colorlinks=false,allbordercolors=blue,pdfborderstyle={/S/U/W 1}}
\usepackage{multirow}  
\usepackage[section]{placeins}

\begin{document}

\markboth{Vandan Patel et. al.}{Spectroscopic study of $ss\bar{b}\bar{b}$ and $bb\bar{b}\bar{b}$ Tetraquarks using Regge phenomenology}

%
\catchline{}{}{}{}{}
%

\title{Spectroscopic study of $ss\bar{b}\bar{b}$ and $bb\bar{b}\bar{b}$ Tetraquarks using Regge phenomenology
}

\author{Vandan Patel
}

\address{Department of Physics, Sardar Vallabhbhai National Institute of Technology, Surat, Gujarat-395007, India\\
	vandankp12998@gmail.com}
	
\author{Juhi Oudichhya}
\address{Department of Physics, Sardar Vallabhbhai National Institute of Technology, Surat, Gujarat-395007, India\\
	juhioudichhya01234@gmail.com}

\author{Ajay Kumar Rai}
\address{Department of Physics, Sardar Vallabhbhai National Institute of Technology, Surat, Gujarat-395007, India\\
	raiajayk@gmail.com}

\maketitle

\begin{history}
\received{Day Month Year}
\revised{Day Month Year}
\accepted{Day Month Year}
\published{Day Month Year}
\end{history}

\begin{abstract}
In this work, we investigate the mass spectra of doubly strange–doubly bottom ($ss\bar{b}\bar{b}$) and fully bottom ($bb\bar{b}\bar{b}$) tetraquark states using the framework of Regge phenomenology. By modeling these tetraquarks as bound diquark–antidiquark systems, we employ quasi-linear Regge trajectories to study both orbital and radial excitations. Within this formalism, we derive linear and quadratic mass inequalities that impose constraints on the masses of ground and excited states of tetraquarks in the $(J, M^2)$ plane. We further extend the analysis to radial excitations through the construction of Regge trajectories in the $(n, M^2)$ plane. Our results show that this phenomenological approach offers a simple yet robust tool for describing tetraquark mass spectra, reproducing trends consistent with other theoretical predictions. The findings serve as valuable references for future experimental searches and may aid in the spin–parity classification of exotic hadrons. Overall, this study contributes to a deeper understanding of multiquark dynamics and hadron spectroscopy within the framework of Quantum Chromodynamics.
\end{abstract}


\ccode{PACS numbers: 03.65.$-$w, 04.62.+v}

\section{Introduction}	
The origins of hadron spectroscopy trace back to the discovery of the pion in 1947 and the subsequent formulation of the quark model during the 1960s~\cite{Lattes1947}. Early experimental findings identified mesons as quark–antiquark ($q\bar{q}$) bound states and baryons as systems of three quarks ($qqq$), leading Gell-Mann and Zweig to propose the quark model in 1964~\cite{ref1,Zweig:1964jf}.

In recent years, numerous hadronic bound states have been detected across several experimental platforms, such as LHCb~\cite{Sigma_b(6097),Cascade_b(6227),Omega_b2020,Cascade_b(6333)}, Belle~\cite{Belle2010,ref4}, BESIII~\cite{BESIII2019,BESIII2020,Exp3,Exp4}, and J-PARC~\cite{K. Aoki2021}. These experimental findings have been complemented by various theoretical models that have successfully predicted their mass spectra and other related properties~\cite{ref5,ref6,ref9}.

While Quantum Chromodynamics (QCD) is the underlying theory that describes strong interactions, it does not restrict the hadronic spectrum solely to conventional mesons and baryons. Instead, QCD allows for the existence of more exotic configurations, such as tetraquarks ($qq\bar{q}\bar{q}$), pentaquarks ($qqqq\bar{q}$), hybrid mesons (comprising a quark--antiquark pair coupled with excited gluonic fields), and glueballs (bound states made entirely of gluons). Initially regarded as purely theoretical constructs, these exotic hadrons have received increasing experimental support over the past two decades.

Significant advancements have been achieved in the identification of exotic hadrons, particularly tetraquark and pentaquark candidates, since the observation of the $X(3872)$ state by the Belle Collaboration in 2003~\cite{Choi:2003ue}. Following this discovery, several other unconventional hadronic states have been reported, including $T_{c\bar{c}1}(3900)$~\cite{BESIII:2020oph,pdg}, $X(4140)$~\cite{PhysRevLett.118.022003,pdg}, , $T_{b\bar{b}1}(10610)$~\cite{Belle:2013urd,pdg} as well as the pentaquark states $P_c(4380)^+$ and $P_c(4450)^+$ observed by the LHCb Collaboration~\cite{Aaij:2015tga}.

In 2009, the CDF Collaboration announced the observation of the $X(4140)$ resonance, with a measured mass of $M = 4143.0 \pm 2.9 \pm 1.2$~MeV and a width of $\Gamma = 11.7^{+8.4}_{-6.7} \pm 3.7$~MeV, in the $B^+ \rightarrow J/\psi \phi K^+$ decay channel~\cite{Aaltonen:2009tz}. In the subsequent years, another nearby structure, the $X(4100)$, was reported by several collaborations, including LHCb, D$\phi$, CMS, and BABAR~\cite{Abazov:2018bun,Sirunyan:2020qir,Aaij:2016iza}. Furthermore, the $X(4274)$ state was detected in 2011 by the CDF Collaboration with a mass of $M = 4274.4 \pm 1.9$~MeV and a width of $\Gamma = 32.3 \pm 7.6$~MeV in the same decay channel, exhibiting a significance of $3.1\sigma$~\cite{CDF:2017hse}. The LHCb Collaboration later confirmed both the $X(4140)$ and $X(4274)$ states and identified their quantum numbers as $J^{PC} = 1^{++}$~\cite{PhysRevLett.118.022003,LHCb:2016nsl}. 

These discoveries have significantly advanced the study of exotic hadrons, particularly tetraquark candidates, and have sparked considerable interest from both experimental and theoretical perspectives. More recently, investigations into properties such as mass spectra and decay behavior of tetraquark systems have also been conducted~\cite{Lodha:2024light}.

In the context of QCD, tetraquark states occupy a natural place: since quarks carry color, any color-singlet combination of quarks is in principle allowed. In ordinary mesons and baryons the color charges of the constituents sum to a singlet, and analogously a tetraquark can be formed by binding two quarks in a color–antitriplet diquark and two antiquarks in a color triplet antidiquark, or by molecular binding of two mesons. Although the detailed dynamics of such multiquark configurations is complex, their existence does not violate QCD.

The investigation of mass spectra of exotic hadrons, such as tetraquarks, plays a vital role in deepening our understanding of the strong interaction as described by Quantum Chromodynamics (QCD)~\cite{ref28}. These studies are particularly important for probing the non-perturbative regime of QCD, which is essential for a comprehensive description of hadronic matter. Additionally, the identification and analysis of tetraquark states provide valuable insights into the mechanisms of color confinement and the role of color charge within QCD~\cite{ref29}.

A wide array of theoretical approaches has been employed to explore the properties of tetraquarks. These include first-principles techniques like lattice QCD, as well as methods such as QCD sum rules, effective field theories, and phenomenological models like the quark model and the diquark–antidiquark model. Among these, lattice QCD is particularly noteworthy for its rigorous foundation, using discretized spacetime grids to simulate QCD and predict properties of multiquark systems~\cite{ref36}. On the other hand, phenomenological approaches offer intuitive interpretations and can be calibrated with experimental data to reliably estimate tetraquark mass spectra~\cite{ref38}. Within the framework of the quark model, tetraquarks are treated as bound states of quarks and antiquarks, analogous to mesons and baryons. The interactions among the constituent quarks are often modeled using effective potentials, such as the widely used Cornell potential~\cite{ref59}.

While tetraquark systems composed of heavy and light quarks—such as those containing charm or bottom quarks—have garnered substantial interest, an especially compelling category involves fully heavy configurations, including the all-charm ($cc\bar{c}\bar{c}$) and all-bottom ($bb\bar{b}\bar{b}$) tetraquarks. These systems have been thoroughly examined through a range of theoretical approaches, such as potential models~\cite{Debastiani:2017msn,Bai:2016int}, QCD sum rules~\cite{Chen:2016jxd}, lattice QCD calculations~\cite{Francis:2016hui}, and the diquark–antidiquark framework~\cite{Karliner:2016zzc}.

In addition to the fully heavy sector, the study of heavy--strange tetraquarks, especially those comprising two strange and two bottom quarks, represents another compelling area of research. These systems exhibit a distinctive interplay between heavy-quark dynamics and SU(3) flavor symmetry breaking, making them ideal candidates for probing the structure of exotic hadrons. Several theoretical studies have explored the properties of doubly strange–doubly bottom tetraquark states using diverse methodologies~\cite{Song2023, Braaten, Ebert2007, Wen2021, Luo2017, PhysRevD.102.034012}. Identifying such states in the experimental spectrum and associating them with observed resonances could offer valuable insights into the underlying mechanisms of multiquark interactions.

In this work, we examine the mass spectra of the $ss\bar{b}\bar{b}$ and $bb\bar{b}\bar{b}$ tetraquark systems using the framework of Regge phenomenology. Building on the quasi-linear Regge trajectory formalism, Wei et al.~\cite{ref39,Wei:2016jyk} formulated essential mass relations for hadrons, including quadratic mass equalities and both linear and quadratic mass inequalities. Inspired by their approach, we extend the formalism to derive analogous mass inequalities for excited tetraquark states under the assumption of linear Regge trajectories. Our analysis focuses on exploring the interdependence between Regge slopes, intercepts, and tetraquark masses in both the $(J, M^2)$ and $(n, M^2)$ planes. This enables us to predict mass ranges for ground and excited states across various spin–parity configurations.

We investigate the Regge trajectories corresponding to $J^P = 0^+$, $1^+$, and $2^+$ for the $ss\bar{b}\bar{b}$ and $bb\bar{b}\bar{b}$ tetraquark systems to determine their Regge parameters. These extracted parameters are then utilized to compute the mass spectra of both systems. Furthermore, we analyze radial excitations by constructing Regge trajectories in the $(n, M^2)$ plane, allowing us to predict the excited-state masses within this framework. The findings presented here are intended to aid ongoing efforts to identify and classify multiquark states, while also serving as valuable references for future experimental searches.

The structure of this paper is as follows: Section~II outlines the theoretical framework of Regge theory. In Section~III, we evaluate the ground-state mass ranges for the $ss\bar{b}\bar{b}$ tetraquark with quantum numbers $J^P = 0^+$, $1^+$, and $2^+$. We also compute the Regge slopes for these states and predict the mass ranges of their orbitally excited counterparts for both the $ss\bar{b}\bar{b}$ and $bb\bar{b}\bar{b}$ configurations in the $(J, M^2)$ and $(n, M^2)$ planes. Section~IV presents a discussion of our results, and Section~V concludes the study.

\section{Theoretical Framework}
The linear Regge trajectory is a widely adopted phenomenological method in the investigation of hadron spectroscopy. Regge theory provides a comprehensive framework for describing various features of strong interactions, including particle mass spectra, the forces between hadrons, and the high-energy behavior of scattering amplitudes. Multiple theoretical approaches have been introduced to understand Regge trajectories, among which one of the earliest and simplest was proposed by 
Yoichiro Nambu in the 1970s~\cite{ref40,ref41}. He modeled the interaction between a quark and an antiquark as a uniformly stretched flux tube, with light quarks at its ends rotating at the speed of light along a circular path of radius~$R$. The mass generated by this flux tube configuration is expressed as~\cite{ref42}

\begin{equation} \label{eq:1}
	M = 2 \int_{0}^{R} \frac{\sigma}{\sqrt{1 - \nu^2(r)}} \, dr = \pi \sigma R ,
\end{equation}

where $\sigma$ denotes the string tension, which corresponds to the energy (or mass) per unit length of the flux tube. Furthermore, the angular momentum associated with the flux tube is evaluated as

\begin{equation} \label{eq:2}
	J = 2 \int_{0}^{R} \frac{\sigma r \nu(r)}{\sqrt{1 - \nu^2(r)}} \, dr = \frac{\pi \sigma R^2}{2} + c'.
\end{equation}

By employing Eqs.~(\ref{eq:1}) and (\ref{eq:2}), we obtain the following expression.

\begin{equation} \label{eq:3}
	J = \frac{M^2}{2\pi\sigma} + c'',
\end{equation}
where $c'$ and $c''$ are constants of integration. As a result, a linear relationship emerges between the total angular momentum $J$ and the squared mass $M^2$. Chew-Frautschi plots represent hadronic Regge trajectories on the $(J, M^2)$ plane~\cite{ref43}. This framework was applied to study the strong interaction dynamics involving quarks and gluons. Their analysis demonstrated that even the higher excited meson and baryon states, though not yet observed experimentally, tend to align along straight lines in the $(J, M^2)$ plane~\cite{ref43}.

Given that both light and heavy hadrons follow approximately linear Regge trajectories, the most general form of a linear Regge trajectory can be expressed as~\cite{ref39}:

\begin{equation} \label{eq:4}
	J = \beta(M) = \beta(0) + \beta' M^2,
\end{equation}
where $\beta(0)$ denotes the trajectory intercept, and $\beta'$ represents the slope associated with the particle's Regge trajectory.

Using Eq.~(\ref{eq:4}), the following expression for the slope can be derived:

\begin{equation} \label{eq:x}
	\beta'=\frac{(J+1)-J}{M^2_{(J+1)}-M^2_J}
\end{equation}

The Regge slopes and intercepts corresponding to different quark constituents within a meson multiplet characterized by spin-parity $J^P$ (or more precisely, by the quantum numbers $N^{2S+1} L_J$) obey the principles of intercept additivity and inverse slope additivity. These relations were derived using a model based on the topological expansion and the quark–antiquark string picture of hadrons~\cite{ref47} (see also Refs.~\cite{ref39,ref44,ref45,ref46}). These relations are given by, 

\begin{equation} \label{eq:5}
	\beta_{i\bar{i}}(0) + \beta_{j\bar{j}}(0) = 2\beta_{i\bar{j}}(0),
\end{equation}

\begin{equation} \label{eq:6}
	\frac{1}{\beta'_{i\bar{i}}} + \frac{1}{\beta'_{j\bar{j}}} = \frac{2}{\beta'_{i\bar{j}}},
\end{equation}

Here, the indices $i$ and $j$ denote the flavors of the quarks. Equation~(\ref{eq:5}) was initially derived within the framework of the dual resonance model for light quarks~\cite{ref48}. Subsequently, it was shown to be valid in several other approaches, including the dual-analytic model~\cite{ref50}, two-dimensional QCD~\cite{ref49}, and the quark bremsstrahlung model~\cite{ref51}.

In this context, and throughout the following discussion, we consider two-body systems where the quark masses satisfy the condition $m_i \leq m_j$, since Eqs.~(\ref{eq:5}) and~(\ref{eq:6}) are symmetric under the interchange of quark flavors $i$ and $j$.

\subsection{Correlation between slope ratios and particle masses
}

For two-body systems, by solving Eqs.~(\ref{eq:4}) and~(\ref{eq:5}) simultaneously, the intercepts can be eliminated, leading to the following expression:

\begin{equation} \label{eq:7}
	\beta'_{i\bar{i}} M_{i\bar{i}}^2 + \beta'_{j\bar{j}} M_{j\bar{j}}^2 = 2 \beta'_{i\bar{j}} M_{i\bar{j}}^2,
\end{equation}
Combining Eqs.~(\ref{eq:6}) and~(\ref{eq:7}) yields two sets of solutions, formulated in terms of slope ratios and hadron masses, as given below:

\begin{equation} \label{eq:9}
	\frac{\beta'_{j\bar{j}}}{\beta'_{i\bar{i}}} = \frac{1}{2 M_{j\bar{j}}^2} \biggl[ \bigl(4 M_{i\bar{j}}^2 - M_{i\bar{i}}^2 - M_{j\bar{j}}^2 \bigr) \\
	\quad \pm \sqrt{\bigl(4 M_{i\bar{j}}^2 - M_{i\bar{i}}^2 - M_{j\bar{j}}^2 \bigr)^2 - 4 M_{i\bar{i}}^2 M_{j\bar{j}}^2} \biggr],
\end{equation}

and

\begin{equation} \label{eq:10}
	\frac{\beta'_{i\bar{j}}}{\beta'_{i\bar{i}}} = \frac{1}{4 M_{i\bar{j}}^2} \biggl[ \bigl(4 M_{i\bar{j}}^2 + M_{i\bar{i}}^2 - M_{j\bar{j}}^2 \bigr) \\
	\quad \pm \sqrt{\bigl(4 M_{i\bar{j}}^2 - M_{i\bar{i}}^2 - M_{j\bar{j}}^2 \bigr)^2 - 4 M_{i\bar{i}}^2 M_{j\bar{j}}^2} \biggr].
\end{equation}	


In this study, we adopt the solutions containing the positive sign in front of the square root term, as they produce slope ratios that are in better agreement with experimentally observed values for certain well-established meson multiplets~\cite{ref39}. Similarly, for tetraquark systems, when computing slope ratios using Eq.~(\ref{eq:x}), the results are more consistent with those derived from the solution involving the plus sign, as opposed to the minus sign. This consistency is verified by comparing the ratio of $\beta'_{cc\bar{c}\bar{c}}$ to $\beta'_{bb\bar{b}\bar{b}}$, using the theoretical mass inputs from Ref.~\cite{ref1050}. 

Thus, the expressions associated with the solutions containing the positive square root sign can be formulated as follows, providing meaningful relationships between the slope ratios and the masses of two-body systems,

\begin{equation} \label{eq:100}
	\frac{\beta'_{j\bar{j}}}{\beta'_{i\bar{i}}} = \frac{1}{2 M_{j\bar{j}}^2} \biggl[ \bigl(4 M_{i\bar{j}}^2 - M_{i\bar{i}}^2 - M_{j\bar{j}}^2 \bigr) \\
	\quad + \sqrt{\bigl(4 M_{i\bar{j}}^2 - M_{i\bar{i}}^2 - M_{j\bar{j}}^2 \bigr)^2 - 4 M_{i\bar{i}}^2 M_{j\bar{j}}^2} \biggr],
\end{equation}

\begin{equation} \label{eq:200}
	\frac{\beta'_{i\bar{j}}}{\beta'_{i\bar{i}}} = \frac{1}{4 M_{i\bar{j}}^2} \biggl[ \bigl(4 M_{i\bar{j}}^2 + M_{i\bar{i}}^2 - M_{j\bar{j}}^2 \bigr) \\
	\quad + \sqrt{\bigl(4 M_{i\bar{j}}^2 - M_{i\bar{i}}^2 - M_{j\bar{j}}^2 \bigr)^2 - 4 M_{i\bar{i}}^2 M_{j\bar{j}}^2} \biggr].
\end{equation}

\subsection{Inequalities Involving Linear and Quadratic Relations of Masses}

{Equation~(\ref{eq:100}) gives rise to two significant inequalities. Since the Regge slopes $\beta'_{j\bar{j}}$ and $\beta'_{i\bar{i}}$ must be positive and real, the ratio $\beta'_{j\bar{j}}/\beta'_{i\bar{i}}$ is also required to be a real quantity. Consequently, from Eq.~(\ref{eq:100}), we obtain:

\begin{equation} \label{eq:15}
	|4M^2_{i\bar{j}} - M^2_{i\bar{i}} - M^2_{j\bar{j}}| \geq 2M_{i\bar{i}}M_{j\bar{j}}.
\end{equation}

When \( i = j \), the condition \( 4M^2_{i\bar{j}} - M^2_{i\bar{i}} - M^2_{j\bar{j}} \leq 0 \) is not fulfilled. Additionally, for \( i \neq j \), this inequality is found to be incompatible with experimental observations from well-known meson multiplets. Hence, we conclude that

\begin{equation}
	4M^2_{i\bar{j}} - M^2_{i\bar{i}} - M^2_{j\bar{j}} \geq 0.
\end{equation}  
As a result, Eq.~(\ref{eq:15}) can be rewritten in the following form:

\begin{equation} \label{eq:16}
	4M^2_{i\bar{j}} - M^2_{i\bar{i}} - M^2_{j\bar{j}} \geq 2M_{i\bar{i}}M_{j\bar{j}}.
\end{equation}

From the above equation we can have:

\begin{equation} \label{eq:17}
	2M_{i\bar{j}} \geq M_{i\bar{i}} + M_{j\bar{j}}.
\end{equation}

For the special case where \( i = j \), it directly follows that \( M_{i\bar{i}} = M_{i\bar{j}} = M_{j\bar{j}} \), leading to the relation \( 2M_{i\bar{j}} = M_{i\bar{i}} + M_{j\bar{j}} \).

Conversely, even when \( i \neq j \), if the condition \( 2M_{i\bar{j}} = M_{i\bar{i}} + M_{j\bar{j}} \) holds, then Eq.~(\ref{eq:100}) enables us to derive the following result:

\begin{equation} \label{eq:18}
	\frac{\beta'_{j\bar{j}}}{\beta'_{i\bar{i}}} = \frac{M_{i\bar{i}}}{M_{j\bar{j}}}.
\end{equation}

The derivation of Eq.~(\ref{eq:18}) clearly demonstrates its applicability to mesons within the same multiplet. Given that hadrons residing on the same Regge trajectory share an identical slope, we obtain:

\begin{equation} \label{eq:19}
	\frac{\beta'_{j\bar{j}}}{\beta'_{i\bar{i}}} = \frac{M_{i\bar{i},J}}{M_{j\bar{j},J}} = \frac{M_{i\bar{i},J+2}}{M_{j\bar{j},J+2}}.
\end{equation}

Using Eq.~(\ref{eq:x}), the slopes of specific Regge trajectories can be determined. For mesons consisting of $i\bar{i}$ and $j\bar{j}$ quark pairs, the corresponding slopes are given by:

\begin{equation} \label{eq:20}
	\beta'_{i\bar{i}} = \frac{2}{M^2_{i\bar{i},J+2} - M^2_{i\bar{i},J}}, \quad \beta'_{j\bar{j}} = \frac{2}{M^2_{j\bar{j},J+2} - M^2_{j\bar{j},J}}. 
\end{equation}

So, by above equation we get,

\begin{equation} \label{eq:21}
	\frac{\beta'_{j\bar{j}}}{\beta'_{i\bar{i}}} = \frac{M_{i\bar{i},J+2} + M_{i\bar{i},J}}{M_{j\bar{j},J+2} + M_{j\bar{j},J}} \times \frac{M_{i\bar{i},J+2} - M_{i\bar{i},J}}{M_{j\bar{j},J+2} - M_{j\bar{j},J}}. 
\end{equation}

The combination of Eqs.~(\ref{eq:19}) and~(\ref{eq:21}) leads to:

\begin{equation} \label{eq:22}
	\frac{\beta'_{j\bar{j}}}{\beta'_{i\bar{i}}} = \frac{M_{i\bar{i},J+2} + M_{i\bar{i},J}}{M_{j\bar{j},J+2} + M_{j\bar{j},J}} \times \frac{M_{i\bar{i},J+2} - M_{i\bar{i},J}}{M_{j\bar{j},J+2} - M_{j\bar{j},J}} = \left( \frac{\beta'_{j\bar{j}}}{\beta'_{i\bar{i}}} \right)^2. 
\end{equation}

As previously noted, the Regge slope $\beta'$ must be a positive real value. Accordingly, from Eq.~(\ref{eq:22}), the condition $\beta'_{j\bar{j}}/\beta'_{i\bar{i}} = 1$ is satisfied when $2M_{i\bar{j}} = M_{i\bar{i}} + M_{j\bar{j}}$. This, in turn, implies through Eq.~(\ref{eq:19}) that $M_{i\bar{i},J} = M_{j\bar{j},J}$ and $M_{i\bar{i},J+2} = M_{j\bar{j},J+2}$, suggesting that $i = j$, provided both $i\bar{i}$ and $j\bar{j}$ meson states share the same $J^P$ quantum numbers.

Based on the preceding analysis, it can be concluded that the relation $2M_{i\bar{j}} = M_{i\bar{i}} + M_{j\bar{j}}$ holds true only in the case where $i = j$. Hence, for $i \neq j$, Eq.~(\ref{eq:17}) yields:

\begin{equation} \label{eq:23}
	2M_{i\bar{j}} > M_{i\bar{i}} + M_{j\bar{j}}. 
\end{equation}

By the equation we can have the following relation:
\begin{equation} \label{eq:24}
	M_{i\bar{j}} > \frac{M_{i\bar{i}} + M_{j\bar{j}}}{2}. 
\end{equation}

Studies have shown that Regge trajectory slopes generally decrease with increasing quark mass~\cite{ref47,ref44,ref45,ref100,ref101,ref102,ref103,ref104,ref53}. As a result, when the mass of quark $j$ is greater than that of quark $i$, the ratio $\beta'_{j\bar{j}}/\beta'_{i\bar{i}}$ becomes less than one. Therefore, using Eq.~(\ref{eq:100}), we obtain:

\begin{equation} \label{eq:25}
	\frac{1}{2M_{j\bar{j}}^{2}} \Bigg[ \left( 4M_{i\bar{j}}^{2} - M_{i\bar{i}}^{2} - M_{j\bar{j}}^{2} \right) \\
	+ \sqrt{\left( 4M_{i\bar{j}}^{2} - M_{i\bar{i}}^{2} - M_{j\bar{j}}^{2} \right)^{2} - 4M_{i\bar{i}}^{2} M_{j\bar{j}}^{2}} \Bigg] < 1
\end{equation}

Since the square root term in the above expression is positive, it follows that:

\begin{equation} \label{eq:26}
	2M_{j\bar{j}}^2 - (4M_{i\bar{j}}^2 - M_{i\bar{i}}^2 - M_{j\bar{j}}^2) > 0
\end{equation}

Using Eqs. (\ref{eq:25}) and (\ref{eq:26}) we get,	

\begin{equation} \label{eq:27}
	(4M_{i\bar{j}}^2 - M_{i\bar{i}}^2 - M_{j\bar{j}}^2)^2 
	- 4M_{i\bar{i}}^2 M_{j\bar{j}}^2 \\
	< \left[ 2M_{j\bar{j}}^2 - (4M_{i\bar{j}}^2 - M_{i\bar{i}}^2 - M_{j\bar{j}}^2) \right]^2
\end{equation}

The preceding two equations can be combined to obtain the following relation:

\begin{equation} \label{eq:28}
	2M_{i\bar{j}}^2 < M_{i\bar{i}}^2 + M_{j\bar{j}}^2
\end{equation}

So, we get,
\begin{equation} \label{eq:29}
	M_{i\bar{j}} < \sqrt{\frac{M_{i\bar{i}}^2 + M_{j\bar{j}}^2}{2}}
\end{equation}

Applying Eqs.~(\ref{eq:24}) and~(\ref{eq:29}) yields the following constraint relation for:
$M_{i\bar{j}}$.

\begin{equation} \label{eq:30}
	\frac{M_{i\bar{i}} + M_{j\bar{j}}}{2} < M_{i\bar{j}} < \sqrt{\frac{M_{i\bar{i}}^2 + M_{j\bar{j}}^2}{2}},
\end{equation}

The mass inequality presented above defines both the upper and lower bounds for the mass of the $M_{i\bar{j}}$ meson. In the following section, this relation will be employed to estimate the mass ranges of yet-to-be-observed tetraquark states.

\section{Mass Spectra of Tetraquark}
\subsection{The four-quark state in the diquark-antidiquark picture}
In this work, we calculate the mass spectra of doubly strange–doubly bottom ($ss\bar{b}\bar{b}$) and fully bottom ($bb\bar{b}\bar{b}$) tetraquark systems by modeling them as bound states of two constituent clusters: a diquark and an anti-diquark. Each diquark is treated as a pair of quarks coupled without any internal spatial excitation. Since a quark pair cannot form a color singlet on its own, a diquark is regarded as an effective degree of freedom observable only within hadrons. A color-singlet tetraquark configuration can arise from two possible diquark–antidiquark combinations: (i) a color anti-triplet diquark with a color triplet anti-diquark $\left( \overline{3} \otimes 3 \right)$, or (ii) a color sextet diquark with a color anti-sextet anti-diquark $\left( 6 \otimes \overline{6} \right)$.

Although Regge phenomenology was originally developed and tested for two-body systems such as mesons and for three-quark baryons, its extension to multiquark states is well motivated when those multiquark states can be approximated as effective two-body systems. In the diquark–antidiquark scheme a compact tetraquark is treated as a bound system of a (compact) diquark and an antidiquark, so that the internal dynamics is dominated by an effective two-body confining interaction. Within this approximation, the fundamental premise of quasi-linear Regge trajectories, confinement between two effective color sources, continues to be valid.

Many works that model tetraquarks explicitly in the diquark–antidiquark picture arrive at the same effective two-body viewpoint and compute spectra using methods very similar to those used for mesons; representative examples include the relativistic quark-model studies of Faustov and collaborators\cite{ref167}, the Lippmann–Schwinger treatment while considering tetraquark in diquark-antidiquark picture by Hadizadeh et al.\cite{ref168}, similar treatment by M. Monemzadeh et. al.\cite{ref169},  nonrelativistic diquark–antidiquark analyses by Mutuk\cite{ref170}, and recent calucllations of fully heavy tetraquarks in relativistic diquark–antidiquark description by Galkin and coworkers\cite{ref171}. These investigations justify treating tetraquarks as effective two-body systems in which Regge-type relations may be constructed and tested.

In particular, a recent dedicated study \cite{ref172} explicitly constructing Regge trajectories for heavy tetraquarks finds approximately linear $(n,M^2)$ trajectories for several heavy tetraquark families including all heavy and doubly heavy-doubly light tetraquarks, supporting the use of Regge phenomenology in this context.

By treating the tetraquark as an effective two-body system comprising a diquark and an anti-diquark, we can utilize Eq.~(\ref{eq:30}) to estimate the corresponding mass ranges.

\subsection{Analysis of $ss\bar{b}\bar{b}$ and $bb\bar{b}\bar{b}$ Tetraquark Mass Spectra in the $(J, M^{2})$ Framework}

In this analysis, we apply Eq.~(\ref{eq:30}) to estimate the mass range of the ground state of the $ss\bar{b}\bar{b}$ tetraquark. This tetraquark is modeled as a bound state of an $ss$ diquark and a $\bar{b}\bar{b}$ anti-diquark, where $s$ and $b$ denote the strange and bottom quarks, respectively. Substituting $i = [ss]$ and $j = [bb]$ into Eq.~(\ref{eq:30}) yields the following relation:

\begin{equation} \label{eq:31}
	\frac{M_{ss\bar{s}\bar{s}} + M_{bb\bar{b}\bar{b}}}{2} < M_{ss\bar{b}\bar{b}} < \sqrt{\frac{M_{ss\bar{s}\bar{s}}^2 + M_{bb\bar{b}\bar{b}}^2}{2}},
\end{equation}

In this work, we use the masses of the $ss\bar{s}\bar{s}$ and $bb\bar{b}\bar{b}$ states from Refs.~\cite{ref55} and~\cite{Liu:2019zuc}, respectively, as theoretical inputs, given the lack of experimental data. By incorporating the theoretical masses of $ss\bar{s}\bar{s}$ and $bb\bar{b}\bar{b}$ tetraquarks with quantum numbers $J^{PC} = 0^{++}$, $1^{+-}$, and $2^{++}$ into our formalism, we estimate the ground-state mass ranges of the $ss\bar{b}\bar{b}$ tetraquark to be 10.808--13.758~GeV for $0^{++}$, 10.826--13.766~GeV for $1^{+-}$, and 10.860--13.779~GeV for $2^{++}$.

To estimate the masses of higher excited states, we calculate the Regge slopes associated with the $ss\bar{b}\bar{b}$ tetraquark system. In particular, the slope parameter $\beta'$ for the $ss\bar{b}\bar{b}$ configuration is determined using Eq.~(\ref{eq:200}). By substituting appropriate values for $i$ and $j$ into the equation and solving for $\beta'_{ss\bar{b}\bar{b}}$, we obtain the following expression:

\begin{equation} \label{eq:32}
	\beta'_{ss\bar{b}\bar{b}} = \frac{\beta'_{ss\bar{s}\bar{s}}}{4 M_{ss\bar{b}\bar{b}}^2} \biggl[ \bigl(4 M_{ss\bar{b}\bar{b}}^2 + M_{ss\bar{s}\bar{s}}^2 - M_{bb\bar{b}\bar{b}}^2 \bigr) \\
	\quad + \sqrt{\bigl(4 M_{ss\bar{b}\bar{b}}^2 - M_{ss\bar{s}\bar{s}}^2 - M_{bb\bar{b}\bar{b}}^2 \bigr)^2 - 4 M_{ss\bar{s}\bar{s}}^2 M_{bb\bar{b}\bar{b}}^2} \biggr]
\end{equation}

The slope of the Regge trajectory for the \(ss\bar{s}\bar{s}\) tetraquark can be determined using equation (\ref{eq:x}).

\begin{equation} \label{eq:33}
	\beta'_{ss\bar{s}\bar{s}} = \frac{1}{M_{ss\bar{s}\bar{s}(1^-)}^2 - M_{ss\bar{s}\bar{s}(0^+)}^2} .
\end{equation} 

The mass of the \(ss\bar{s}\bar{s}\) tetraquark with \(J^P = 1^-\), as reported in Ref.~\cite{ref55}, is employed to determine its corresponding Regge slope. The computed slope values for various \(J^P\) quantum numbers are summarized in Table~\ref{tab:Slope}.

Substituting the values of \(M_{ss\bar{s}\bar{s}}\), \(M_{bb\bar{b}\bar{b}}\), and \(\beta'_{ss\bar{s}\bar{s}}\) into Eq.~(\ref{eq:32}), the slope \(\beta'_{ss\bar{b}\bar{b}}\) can be expressed as a function of the mass \(M_{ss\bar{b}\bar{b}}\). This function exhibits an increasing trend within the mass interval 10.808--13.758~GeV. For the case \(J^P = 0^+\), the estimated range of \(\beta'_{ss\bar{b}\bar{b}}\) spans from 0.12549 to 0.59149, as listed in Table~\ref{tab:Slope}. The slope ranges corresponding to other \(J^P\) quantum numbers are also presented in the same table.

By combining Eqs.~(\ref{eq:6}) and~(\ref{eq:32}), we arrive at the following expression:
\\
\\

\begin{equation} \label{eq:101}
	\beta'_{bb\bar{b}\bar{b}} = \frac{1}{
		\left(
		\frac{2}{\beta'_{ss\bar{s}\bar{s}} \cdot \frac{1}{4 M_{ss\bar{b}\bar{b}}^{2}} \left( \left(4 M_{ss\bar{b}\bar{b}}^{2} + M_{ss\bar{s}\bar{s}}^{2} - M_{bb\bar{b}\bar{b}}^{2} \right) + \sqrt{ \left(4 M_{ss\bar{b}\bar{b}}^{2} - M_{ss\bar{s}\bar{s}}^{2} - M_{bb\bar{b}\bar{b}}^{2} \right)^{2} - 4 M_{ss\bar{s}\bar{s}}^{2} M_{bb\bar{b}\bar{b}}^{2} } \right)}-\frac{1}{\beta'_{ss\bar{s}\bar{s}}}
		\right)
	}
\end{equation}

By substituting the values of \(M_{ss\bar{s}\bar{s}}\), \(M_{bb\bar{b}\bar{b}}\), and \(\beta'_{ss\bar{s}\bar{s}}\) into the above equation, the slope \(\beta'_{bb\bar{b}\bar{b}}\) can be expressed as a function of \(M_{ss\bar{b}\bar{b}}\). Over the range 10.808--13.758~GeV, this function shows an increasing behavior. For \(J^P = 0^+\), the estimated range of \(\beta'_{bb\bar{b}\bar{b}}\) lies between 0.07019 and 0.59149, as detailed in Table~\ref{tab:Slope}. The slope intervals for other \(J^P\) configurations are also included in the same table.

Moreover, applying Eq.~(\ref{eq:x}), the mass of the excited $ss\bar{b}\bar{b}$ tetraquark state can be written as:

\begin{equation} \label{eq:34}
	M_{J+k(ss\bar{b}\bar{b})} = \sqrt{M_{J(ss\bar{b}\bar{b})}^2 + \frac{k}{\beta'_{ss\bar{b}\bar{b}}}} ,
\end{equation}
here, k is an positive integer number.

Using Eqs.~(\ref{eq:34}) and~(\ref{eq:32}), the mass \( M_{J+k(ss\bar{b}\bar{b})} \) can be expressed in terms of the ground-state mass \( M_{J(ss\bar{b}\bar{b})} \). 

As an example, by setting \(J^P = 0^+\) and \(k = 1\) in Eq.~(\ref{eq:34}), the resulting mass range for the $ss\bar{b}\bar{b}$ tetraquark with \(J^P = 1^-\) falls between 11.131~GeV and 13.820~GeV, corresponding to the interval (10.808--13.758) for the ground state. This range is listed in Table~\ref{table:ssbb_tetraquarks}. Similarly, the masses of other excited states of the \( ss\bar{b}\bar{b} \) tetraquark are computed and reported in Table~\ref{table:ssbb_tetraquarks}. Comparisons with predictions from other theoretical studies are provided in Table 3 for the \(ss\bar{b}\bar{b}\) system.

\begin{table}[ph]
	\tbl{Values of Regge slopes for \(ss\bar{s}\bar{s}\), \(bb\bar{b}\bar{b}\), and \(ss\bar{b}\bar{b}\) tetraquarks in the \((J, M^2)\) plane (in \(\text{GeV}^{-2}\)).\label{tab:Slope}}
	{\begin{tabular}{@{}cccc@{}} \toprule
			\(J^P\) & \(\beta'_{ss\bar{s}\bar{s}} \, (\text{GeV}^{-2})\) & \(\beta'_{ss\bar{b}\bar{b}} \, (\text{GeV}^{-2})\) & \(\beta'_{bb\bar{b}\bar{b}} \, (\text{GeV}^{-2})\) \\
			\colrule
			\(0^+\) & 0.59149 & 0.12549--0.59149 & 0.07019--0.59149 \\
			\(1^+\) & 0.57189 & 0.12271--0.57189 & 0.06873--0.57189 \\
			\(2^+\) & 0.57535 & 0.12599--0.57535 & 0.07074--0.57535 \\
			\botrule
	\end{tabular}}
\end{table}

Similarly, we can get the corresponding formula for the \(bb\bar{b}\bar{b}\) tetraquark using Eq. (\ref{eq:x}).

\begin{equation} \label{eq:35}
	M_{J+k(bb\bar{b}\bar{b})} = \sqrt{M_{J(bb\bar{b}\bar{b})}^2 + \frac{k}{\beta'_{bb\bar{b}\bar{b}}}} ,
\end{equation}

By applying Eqs.~(\ref{eq:35}) and~(\ref{eq:101}), the mass \( M_{J+k(bb\bar{b}\bar{b})} \) can be expressed as a function of \( M_{J(ss\bar{b}\bar{b})} \). Using this approach, we have calculated the mass ranges for the excited states of the \( bb\bar{b}\bar{b} \) tetraquark, as summarized in Table~\ref{table:all_bottom_tetraquarks}.

As discussed earlier, the predicted mass spectra for the \(ss\bar{b}\bar{b}\) and \(bb\bar{b}\bar{b}\) tetraquark systems are presented in Tables~\ref{table:ssbb_tetraquarks} and~\ref{table:all_bottom_tetraquarks}, respectively. These results are also compared against the relevant two-meson threshold values. Additionally, comparisons with predictions from other theoretical studies are provided in Table~\ref{table:all_bottom_tetraquarks_comparison} for the all-bottom tetraquark states.

In addition, we have illustrated the Regge trajectories of the $ss\bar{b}\bar{b}$ and $bb\bar{b}\bar{b}$ tetraquarks in the $(J,M^2)$ plane, shown in Figures~\ref{fig:regge trajectory for ssbb of S=0 in J plane} and \ref{fig:regge trajectory for bbbb of S=0 in J plane} for \(S=0\). Each figure presents two trajectories: one corresponding to the lower mass bounds and the other to the upper mass bounds.

\begin{table}[ph]
	\tbl{Mass spectra of $ss\bar{b}\bar{b}$ tetraquarks with their corresponding two-meson thresholds.\label{table:ssbb_tetraquarks}}
	{\begin{tabular}{@{}ccccc@{}} \toprule
			State & $J^{PC}$ & Calculated Mass (GeV) & Two-meson threshold & Threshold Mass (GeV) \\
			\colrule
			$1^1S_0$  & $0^{++}$ & 10.808--13.758 & $B^0_sB^0_s$ & 10.734 \\
			$1^1P_1$  & $1^{--}$ & 11.130--13.820 & $B^0_sB_{s1}(5830)^0$ & 11.196 \\
			$1^1D_2$  & $2^{++}$ & 11.394--13.881 & $B^*_sB^*_s$ & 10.830 \\
			$1^1F_3$  & $3^{--}$ & 11.623--13.942 & $B^*_sB^*_{s2}(5840)^0$ & 11.255 \\
			$1^1G_4$  & $4^{++}$ & 11.828--14.002 & $B^*_{s2}(5840)^0B^*_{s2}(5840)^0$ & 11.680 \\
			$1^3S_1$  & $1^{+-}$ & 10.826--13.766 & $B^0_sB^*_s$ & 10.782 \\
			$1^3P_2$  & $2^{-+}$ & 11.156--13.829 & $B^*_sB_{s1}(5830)^0$ & 11.244 \\
			$1^3D_3$  & $3^{+-}$ & 11.424--13.892 & $B_{s1}(5830)^0B^*_{s2}(5840)^0$ & 11.669 \\
			$1^3F_4$  & $4^{-+}$ & 11.657--13.955 & -- & -- \\
			$1^3G_5$  & $5^{+-}$ & 11.866--14.018 & -- & -- \\
			$1^5S_2$  & $2^{++}$ & 10.860--13.779 & $B^*_sB^*_s$ & 10.830 \\
			$1^5P_3$  & $3^{--}$ & 11.181--13.842 & $B^*_sB^*_{s2}(5840)^0$ & 11.255 \\
			$1^5D_4$  & $4^{++}$ & 11.446--13.905 & $B^*_{s2}(5840)^0B^*_{s2}(5840)^0$ & 11.680 \\
			$1^5F_5$  & $5^{--}$ & 11.676--13.967 & -- & -- \\
			$1^5G_6$  & $6^{++}$ & 11.882--14.029 & -- & -- \\
			\botrule
	\end{tabular}}
\end{table}

\begin{table}[ph]
	\tbl{Comparison of $ss\bar{b}\bar{b}$ tetraquark masses with previous studies (in GeV).\label{table:ssbb_tetraquarks_comparison}}
	{\begin{tabular}{@{}ccccccccc@{}} \toprule
			State & $J^{PC}$ & This Work & Ref.~\cite{Song2023} & Ref.~\cite{Braaten} & Ref.~\cite{Ebert2007} & Ref.~\cite{Wen2021} & Ref.~\cite{Luo2017} & Ref.~\cite{PhysRevD.102.034012} \\
			\colrule
			$1^1S_0$  & $0^{++}$  & 10.808--13.758 & 10.976 & 10.898 & 10.932 & 11.078 & 11.157 & 10.972 \\
			$1^1P_1$  & $1^{--}$  & 11.130--13.820 & 11.206 & --     & --     & --     & --     & --     \\
			$1^1D_2$  & $2^{++}$  & 11.394--13.881 & --     & --     & --     & --     & --     & --     \\
			$1^1F_3$  & $3^{--}$  & 11.623--13.942 & --     & --     & --     & --     & --     & --     \\
			$1^1G_4$  & $4^{++}$  & 11.828--14.002 & --     & --     & --     & --     & --     & --     \\
			$1^3S_1$  & $1^{+-}$  & 10.826--13.766 & 10.981 & 10.905 & 10.939 & 11.099 & 11.199 & 10.986 \\
			$1^3P_2$  & $2^{-+}$  & 11.156--13.829 & 11.216 & --     & --     & --     & --     & --     \\
			$1^3D_3$  & $3^{+-}$  & 11.424--13.892 & --     & --     & --     & --     & --     & --     \\
			$1^3F_4$  & $4^{-+}$  & 11.657--13.955 & --     & --     & --     & --     & --     & --     \\
			$1^3G_5$  & $5^{+-}$  & 11.866--14.018 & --     & --     & --     & --     & --     & --     \\
			$1^5S_2$  & $2^{++}$  & 10.860--13.779 & 10.991 & 10.919 & 10.950 & 11.119 & 11.224 & 11.004 \\
			$1^5P_3$  & $3^{--}$  & 11.181--13.842 & 11.224 & --     & --     & --     & --     & --     \\
			$1^5D_4$  & $4^{++}$  & 11.446--13.905 & --     & --     & --     & --     & --     & --     \\
			$1^5F_5$  & $5^{--}$  & 11.676--13.967 & --     & --     & --     & --     & --     & --     \\
			$1^5G_6$  & $6^{++}$  & 11.882--14.029 & --     & --     & --     & --     & --     & --     \\
			\botrule
	\end{tabular}}
\end{table}

\begin{table}[ph]
	\tbl{Mass spectra of all bottom tetraquarks with their corresponding two-meson thresholds.\label{table:all_bottom_tetraquarks}}
	{\begin{tabular}{ccccc}
		\toprule
		State & $J^{PC}$ & Calculated & Two-meson & Threshold \\ 
		&       & Mass & threshold & Mass  \\ 
		&       & (GeV) &  & (GeV) \\ 
		\colrule
		$1^1P_1$ & $1^{--}$ & 19.366-19.687 & $\Upsilon(1S)\chi_{b0}(1P) $ & 19.319 \\ 
		$1^1D_2$ & $2^{++}$ & 19.409-20.046 & $\Upsilon(1S)\Upsilon(1S)$ & 18.920 \\ 
		$1^1F_3$ & $3^{--}$ & 19.453-20.398 & $\Upsilon(1S)\chi_{b2}(1P)$ & 19.372 \\ 
		$1^1G_4$ & $4^{++}$ & 19.496-20.744 & $\chi_{b2}(1P)\chi_{b2}(1P)$ & 19.824 \\ 
		$1^3P_2$ & $2^{-+}$ & 19.374-19.702 & $\Upsilon(1S)\chi_{b1}(1P)$ & 19.353 \\ 
		$1^3D_3$ & $3^{+-}$ & 19.419-20.068 & $\Upsilon(1S)\Upsilon_2(1D)$ & 19.624 \\ 
		$1^3F_4$ & $4^{-+}$ & 19.464-20.427 & $\chi_{b2}(1P)\Upsilon_2(1D)$ & 20.076 \\ 
		$1^3G_5$ & $5^{+-}$ & 19.509-20.780 & - & - \\ 
		$1^5P_3$ & $3^{--}$ & 19.386-19.703 & $\Upsilon(1S)\chi_{b2}(1P)$ & 19.372 \\ 
		$1^5D_4$ & $4^{++}$ & 19.431-20.058 & $\chi_{b2}(1P)\chi_{b2}(1P)$ & 19.824 \\ 
		$1^5F_5$ & $5^{--}$ & 19.475-20.408 & - & - \\ 
		$1^5G_6$ & $6^{++}$ & 19.520-20.751 & - & - \\ 
		\botrule
	\end{tabular}}
\end{table}

\begin{table}[ph]
	\tbl{Comparison of all-bottom tetraquark ($bb\bar{b}\bar{b}$) masses with previous studies (in GeV).\label{table:all_bottom_tetraquarks_comparison}}
	{\begin{tabular}{@{}ccccccccccc@{}} \toprule
			State & $J^P$ & This Work & Ref.~\cite{Tiwari2021} & Ref.~\cite{Chen:2024xyz} & Ref.~\cite{PhysRevD.109.076017} & Ref.~\cite{sym14122504} & Ref.~\cite{CHEN2017247} & Ref.~\cite{Wang2019} & Ref.~\cite{Bedolla2020} & Ref.~\cite{PhysRevD.104.116029} \\
			\colrule
			$1^1P_1$  & $1^{--}$ & 19.366--19.687 & 19.361 & $17.650^{+210}_{-190}$ & 19.749/19.795 & 19.536 & $18.770 \pm 0.160$ & $18.890 \pm 0.090$ & 19.281 & -- \\
			$1^1D_2$  & $2^{++}$ & 19.409--20.046 & --     & --                     & --             & 19.715 & --              & --              & 19.510 & 19.669 \\
			$1^1F_3$  & $3^{--}$ & 19.453--20.398 & --     & --                     & --             & --     & --              & --              & --     & --     \\
			$1^1G_4$  & $4^{++}$ & 19.496--20.744 & --     & --                     & --             & --     & --              & --              & --     & --     \\
			$1^3P_2$  & $2^{-+}$ & 19.374--19.702 & 19.373 & $18.230^{+250}_{-240}$ & 19.756/19.609/19.492 & 19.539 & -- & -- & -- & -- \\
			$1^3D_3$  & $3^{+-}$ & 19.419--20.068 & --     & --                     & --             & 19.720 & --              & --              & --     & 19.675 \\
			$1^3F_4$  & $4^{-+}$ & 19.464--20.427 & --     & --                     & --             & --     & --              & --              & --     & --     \\
			$1^3G_5$  & $5^{+-}$ & 19.509--20.780 & --     & --                     & --             & --     & --              & --              & --     & --     \\
			$1^5P_3$  & $3^{--}$ & 19.386--19.703 & 19.388 & $17.620^{+200}_{-180}$ & 19.617         & 19.545 & --              & --              & --     & --     \\
			$1^5D_4$  & $4^{++}$ & 19.431--20.058 & --     & --                     & --             & 19.724 & --              & --              & --     & 19.686 \\
			$1^5F_5$  & $5^{--}$ & 19.475--20.408 & --     & --                     & --             & --     & --              & --              & --     & --     \\
			$1^5G_6$  & $6^{++}$ & 19.520--20.751 & --     & --                     & --             & --     & --              & --              & --     & --     \\
			\botrule
	\end{tabular}}
\end{table}

\begin{figure}[htbp]
	\centering
	\includegraphics[width=\linewidth]{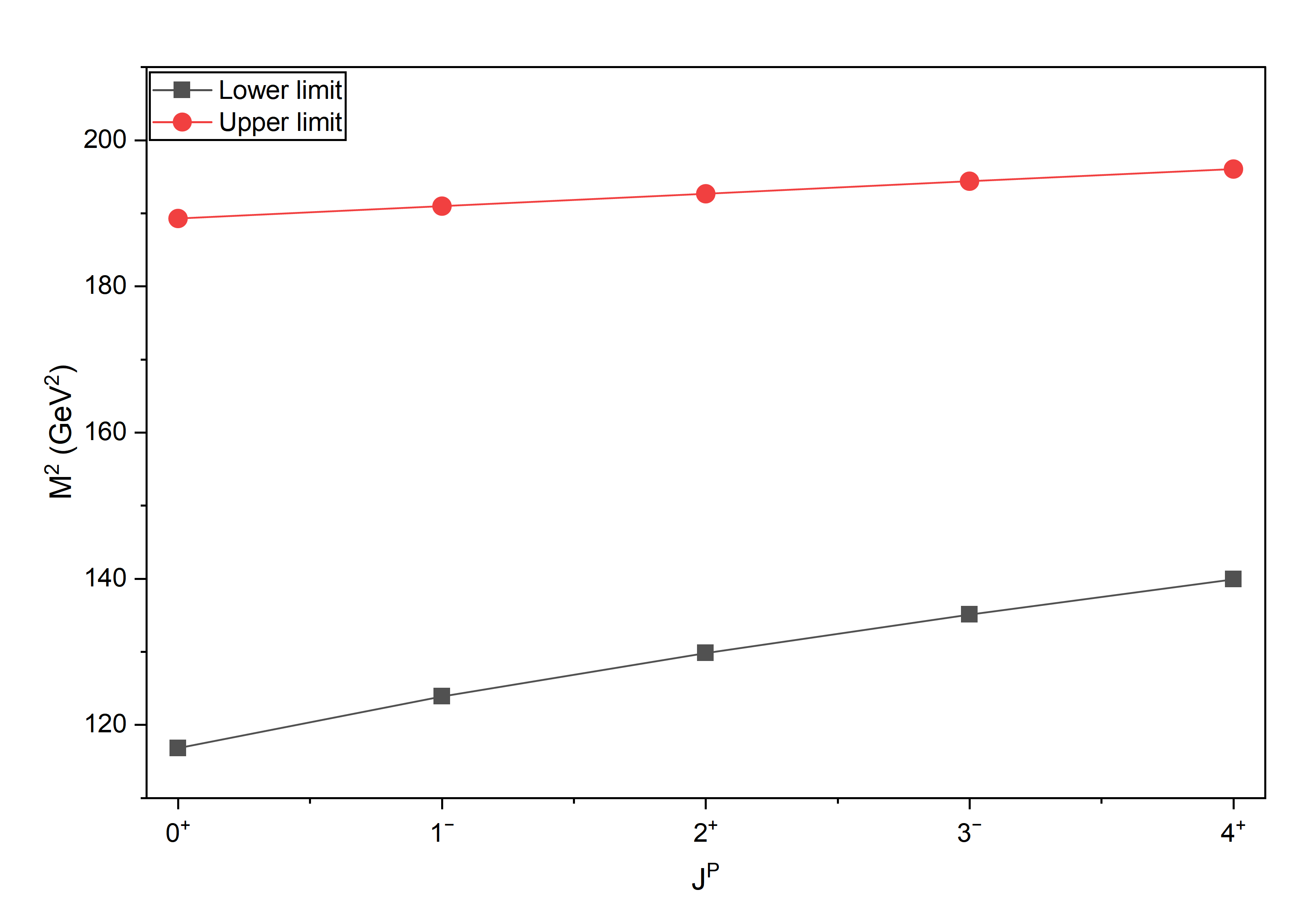}  
	\caption{Regge trajectory of $ss\bar{b}\bar{b}$ tetraqaurak for $S=0$ in $(J,M^2)$ plane.}
	\label{fig:regge trajectory for ssbb of S=0 in J plane}
\end{figure}

\begin{figure}[htbp]
	\centering
	\includegraphics[width=\linewidth]{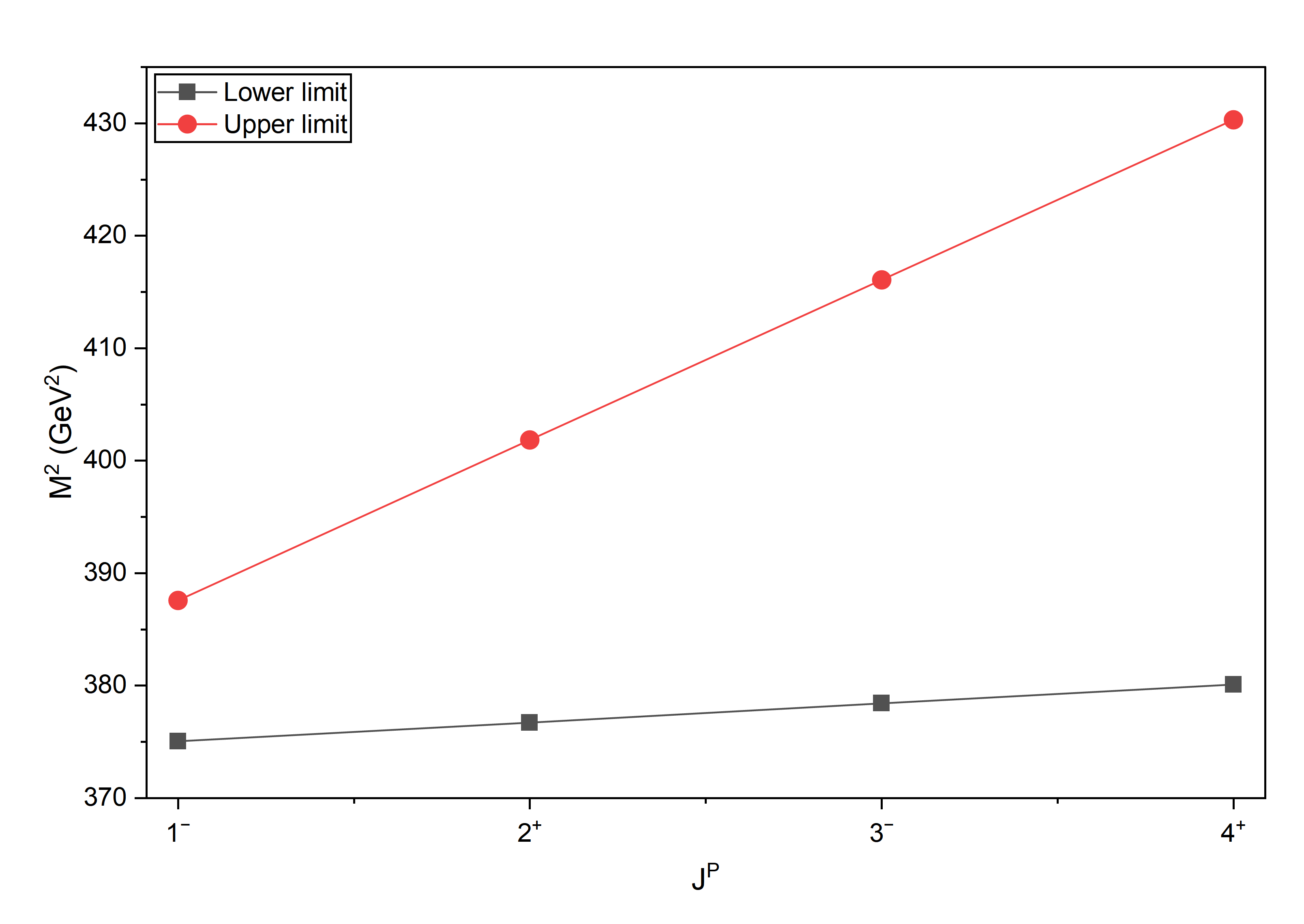}  
	\caption{Regge trajectory of $bb\bar{b}\bar{b}$ tetraqaurak for $S=0$ in $(J,M^2)$ plane.}
	\label{fig:regge trajectory for bbbb of S=0 in J plane}
\end{figure}

\subsection{Analysis of $ss\bar{b}\bar{b}$ and $bb\bar{b}\bar{b}$ Tetraquark Mass Spectra in the $(n, M^{2})$ Framework}

In this section, we compute the Regge parameters for the \(ss\bar{b}\bar{b}\) and \(bb\bar{b}\bar{b}\) tetraquark systems in the \((n, M^2)\) plane to evaluate the masses of their radially excited states. The general linear form of the Regge trajectory in the \((n, M^2)\) plane is expressed as:

\begin{equation} \label{eq:36}
	n = \alpha(M) = \alpha(0) + \alpha' M^2 ,
\end{equation}

Here, \(n = 1, 2, 3, \ldots\) represents the radial quantum number, while \(\alpha(0)\) and \(\alpha'\) denote the intercept and slope of the Regge trajectory in the \((n, M^2)\) plane, respectively. It is assumed that the Regge parameters remain constant for all tetraquark states lying along the same Regge trajectory. In this analysis, we adopt the same method used earlier in the \((J, M^2)\) plane to extract the Regge parameters.

Using Eq.~(\ref{eq:36}), the slope of the Regge trajectory for the $ss\bar{s}\bar{s}$ tetraquark in the \((n, M^2)\) plane can be obtained from the following expression:

\begin{equation} \label{eq:37}
	\alpha'_{ss\bar{s}\bar{s}} = \frac{1}{M_{ss\bar{s}\bar{s}(2S)}^2 - M_{ss\bar{s}\bar{s}(1S)}^2}, 
\end{equation}

In the absence of experimental data, we utilize the theoretically predicted masses of the \( 1^1S_0 \) and \( 2^1S_0 \) states of the fully strange (\( ss\bar{s}\bar{s} \)) tetraquark, as reported in Ref.~\cite{ref55}, for our analysis. Using Eq.~(\ref{eq:37}), we compute the Regge slope as \( \alpha'_{ss\bar{s}\bar{s}} = 0.35544 \, \text{GeV}^{-2} \), which is listed in Table~\ref{tab:SlopenM2} along with the slopes corresponding to other \( J^P \) quantum numbers.

We assume that Eqs.~(\ref{eq:5}) and~(\ref{eq:6}), originally derived for the \((J, M^2)\) plane, remain valid in the \((n, M^2)\) representation as well. Accordingly, by applying Eq.~(\ref{eq:200}), we obtain the following relation within the \((n, M^2)\) framework:

\begin{equation} \label{eq:38}
	\alpha'_{ss\bar{b}\bar{b}} = \frac{\alpha'_{ss\bar{s}\bar{s}}}{4 M_{ss\bar{b}\bar{b}}^2} \biggl[ \bigl(4 M_{ss\bar{b}\bar{b}}^2 + M_{ss\bar{s}\bar{s}}^2 - M_{bb\bar{b}\bar{b}}^2 \bigr) \\
	\quad + \sqrt{\bigl(4 M_{ss\bar{b}\bar{b}}^2 - M_{ss\bar{s}\bar{s}}^2 - M_{bb\bar{b}\bar{b}}^2 \bigr)^2 - 4 M_{ss\bar{s}\bar{s}}^2 M_{bb\bar{b}\bar{b}}^2} \biggr]
\end{equation}

By substituting the ground-state (\(1^1 S_0\)) masses of the \(ss\bar{s}\bar{s}\) and \(bb\bar{b}\bar{b}\) tetraquarks from Refs.~\cite{ref55} and~\cite{Liu:2019zuc}, respectively, along with the slope \(\alpha'_{ss\bar{s}\bar{s}} = 0.35544\), into the above relation, the slope \(\alpha'_{ss\bar{b}\bar{b}}\) can be expressed as a function of \(M_{ss\bar{b}\bar{b}}\). This function exhibits an increasing trend over the interval 10.807--13.758~GeV. The resulting range of \(\alpha'_{ss\bar{b}\bar{b}}\) spans from 0.07541 to 0.35544, as presented in Table~\ref{tab:SlopenM2}.

Following a procedure similar to that employed in the \((J, M^2)\) analysis, the slope parameters for other tetraquark systems—such as the \(bb\bar{b}\bar{b}\) state—have been extracted in the \((n, M^2)\) plane. All slope values obtained within this framework are listed in Table~\ref{tab:SlopenM2}. Additionally, using the same approach adopted for calculating excited-state masses in the \((J, M^2)\) plane, we have evaluated the mass spectra of excited states for the \(ss\bar{b}\bar{b}\) and \(bb\bar{b}\bar{b}\) tetraquarks in the \((n, M^2)\) representation. These results are shown in Tables~\ref{tab:ssbb_spectra_innm^2} and~\ref{tab:bbbb_spectra_innm2}, respectively, along with comparisons to available theoretical predictions.

We have further examined the Regge trajectories in the $(n,M^2)$ plane for the $ss\bar{b}\bar{b}$ and $bb\bar{b}\bar{b}$ tetraquarks, which are depicted in Figures~\ref{fig:regge trajectory for ssbb of S=0 in n plane} and \ref{fig:regge trajectory for bbbb of S=0 in n plane} for \(S=0\). Each figure illustrates two trajectories: one corresponding to the lower mass bounds and the other to the upper mass bounds.

\begin{table}[h]
	\tbl{Values of Regge Slopes for \(ss\bar{s}\bar{s}\), \(ss\bar{b}\bar{b}\) and \(bb\bar{b}\bar{b}\) tetraquarks in \((n,M^2)\) plane (in \(\text{GeV}^{-2}\))\label{tab:SlopenM2}}
	{\begin{tabular}{cccc}
		\toprule
		S & \(\alpha'_{ss\bar{s}\bar{s}} \, (\text{GeV}^{-2})\) & \(\alpha'_{ss\bar{b}\bar{b}} \, (\text{GeV}^{-2})\) & \(\alpha'_{bb\bar{b}\bar{b}} \, (\text{GeV}^{-2})\) \\
		\colrule
		S=0 & 0.35544 & 0.07541-0.35544 & 0.04218-0.35544 \\
		S=1 & 0.35419 & 0.07600-0.35419 & 0.04257-0.35419 \\
		S=2 & 0.31945 & 0.06995-0.31945 & 0.03928-0.31945 \\
		\botrule
	\end{tabular}}
\end{table}

\begin{table}[ph]
	\tbl{Mass spectra of $ss\bar{b}\bar{b}$ tetraquark in $(n, M^2)$ plane (in GeV)\label{tab:ssbb_spectra_innm^2}}
	{\begin{tabular}{cccc}
		\toprule
		Spin & State & $J^P$ & Calculated mass (GeV) \\    
		\colrule
		\multirow{5}{*}{$S=0$}
		& $1^1S_0$ & $0^{++}$ & $10.808$--$13.758$ \\
		& $2^1S_0$ & $0^{++}$ & $11.311$--$13.860$ \\
		& $3^1S_0$ & $0^{++}$ & $11.693$--$13.961$ \\
		& $4^1S_0$ & $0^{++}$ & $12.013$--$14.062$ \\
		& $5^1S_0$ & $0^{++}$ & $12.296$--$14.162$ \\
		\colrule
		\multirow{5}{*}{$S=1$}
		& $1^3S_1$ & $1^{+-}$ & $10.826$--$13.766$ \\
		& $2^3S_1$ & $1^{+-}$ & $11.326$--$13.868$ \\
		& $3^3S_1$ & $1^{+-}$ & $11.707$--$13.970$ \\
		& $4^3S_1$ & $1^{+-}$ & $12.028$--$14.070$ \\
		& $5^3S_1$ & $1^{+-}$ & $12.309$--$14.170$ \\
		\colrule
		\multirow{5}{*}{$S=2$}
		& $1^5S_2$ & $2^{++}$ & $10.860$--$13.779$ \\
		& $2^5S_2$ & $2^{++}$ & $11.396$--$13.892$ \\
		& $3^5S_2$ & $2^{++}$ & $11.802$--$14.004$ \\
		& $4^5S_2$ & $2^{++}$ & $12.142$--$14.116$ \\
		& $5^5S_2$ & $2^{++}$ & $12.441$--$14.226$ \\
		\botrule
	\end{tabular}}
\end{table}

\begin{table}[ph]
	\tbl{Mass spectra of $bb\bar{b}\bar{b}$ tetraquark states in the $(n, M^2)$ plane (in GeV) and comparison with other studies.\label{tab:bbbb_spectra_innm2}}
	{\begin{tabular}{@{}ccccccccccccc@{}} \toprule
			Spin & State & $J^P$ & Calculated Mass & 
			Ref.~\cite{PhysRevD.109.076017} & Ref.~\cite{sym14122504} & Ref.~\cite{Bedolla2020} & Ref.~\cite{Lu2020} & Ref.~\cite{Zhao:2020} & Ref.~\cite{PhysRevD.104.014018} & Ref.~\cite{Tiwari2021} & Ref.~\cite{Mutuk2021} & Ref.~\cite{Ke2021} \\
			\colrule
			\multirow{4}{*}{$S=0$}
			& $2^1S_0$ & $0^{++}$ & 19.395--19.926 & 19.719/19.811 & 19.680 & 19.335 & 19.625 & 19.583 & 19.636 & 19.441 & 19.841 & 19.591 \\
			& $3^1S_0$ & $0^{++}$ & 19.467--20.512 & -- & 19.941 & 19.644 & 19.726 & 19.887 & 19.907 & 19.759 & 20.001 & 19.845 \\
			& $4^1S_0$ & $0^{++}$ & 19.539--21.082 & -- & -- & -- & -- & -- & -- & -- & -- & -- \\
			& $5^1S_0$ & $0^{++}$ & 19.611--21.637 & -- & -- & -- & -- & -- & -- & -- & -- & -- \\
			\colrule
			\multirow{4}{*}{$S=1$}
			& $2^3S_1$ & $1^{+-}$ & 19.402--19.927 & 19.722/19.813 & 19.682 & 19.366 & 19.625 & 19.582 & 19.662 & 19.443 & 19.849 & 19.728 \\
			& $3^3S_1$ & $1^{+-}$ & 19.474--20.508 & -- & 19.943 & 19.665 & 19.733 & 19.889 & 19.930 & 19.760 & 20.012 & 20.016 \\
			& $4^3S_1$ & $1^{+-}$ & 19.547--21.073 & -- & -- & -- & -- & -- & -- & -- & -- & -- \\
			& $5^3S_1$ & $1^{+-}$ & 19.619--21.624 & -- & -- & -- & -- & -- & -- & -- & -- & -- \\
			\colrule
			\multirow{4}{*}{$S=2$}
			& $2^5S_2$ & $2^{++}$ & 19.422--19.988 & 19.726/19.816 & 19.687 & 19.398 & 19.633 & 19.594 & 19.684 & 19.448 & 19.855 & 19.728 \\
			& $3^5S_2$ & $2^{++}$ & 19.502--20.615 & -- & 19.947 & 19.688 & 19.736 & 19.898 & 19.926 & 19.764 & 20.021 & 20.016 \\
			& $4^5S_2$ & $2^{++}$ & 19.582--21.224 & -- & -- & -- & -- & -- & -- & -- & -- & -- \\
			& $5^5S_2$ & $2^{++}$ & 19.662--21.815 & -- & -- & -- & -- & -- & -- & -- & -- & -- \\
			\botrule
	\end{tabular}}
\end{table}

\begin{figure}[htbp]
	\centering
	\includegraphics[width=\linewidth]{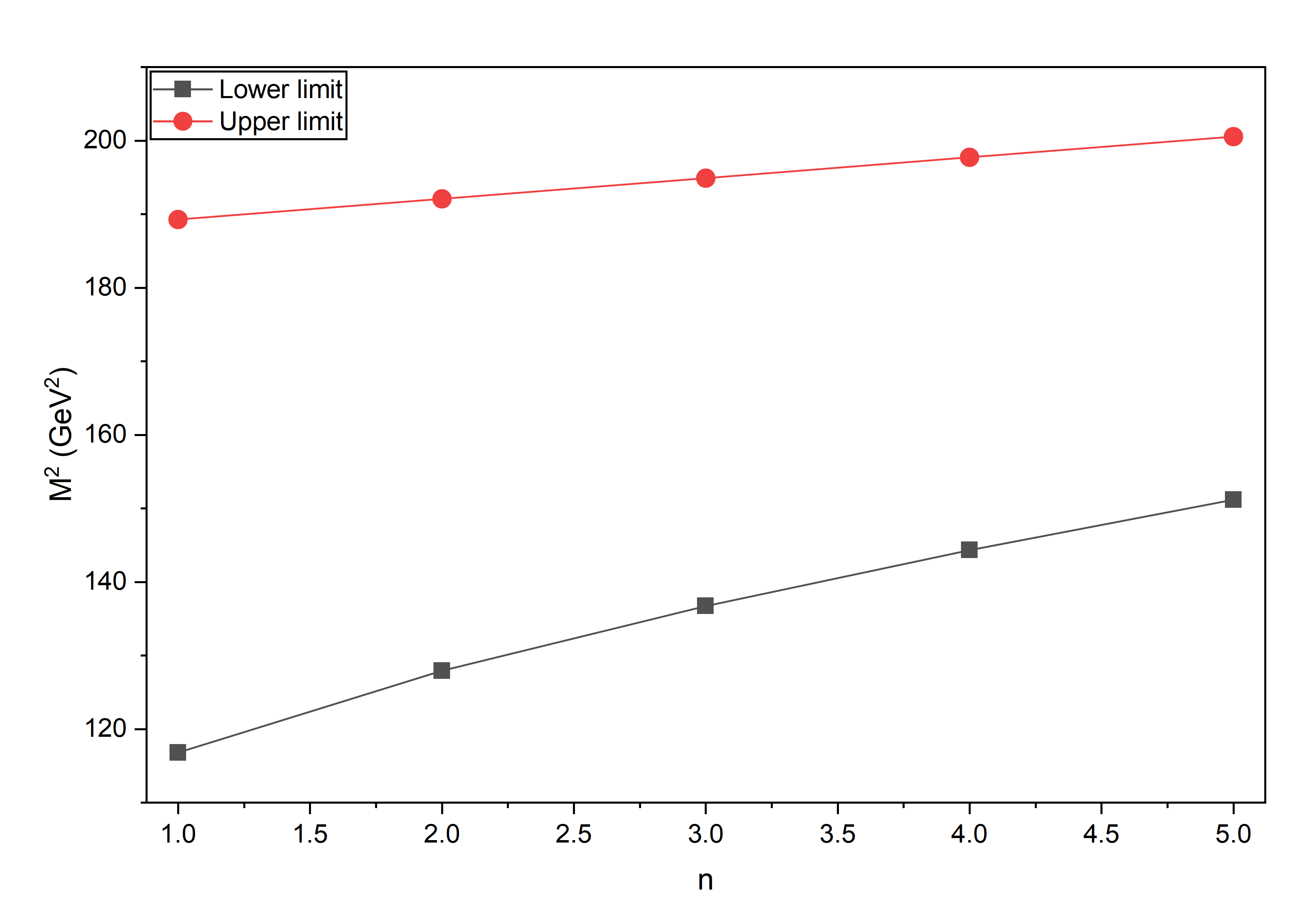}  
	\caption{Regge trajectory of $ss\bar{b}\bar{b}$ tetraqaurak for $S=0$ in $(n,M^2)$ plane.}
	\label{fig:regge trajectory for ssbb of S=0 in n plane}
\end{figure}

\begin{figure}[htbp]
	\centering
	\includegraphics[width=\linewidth]{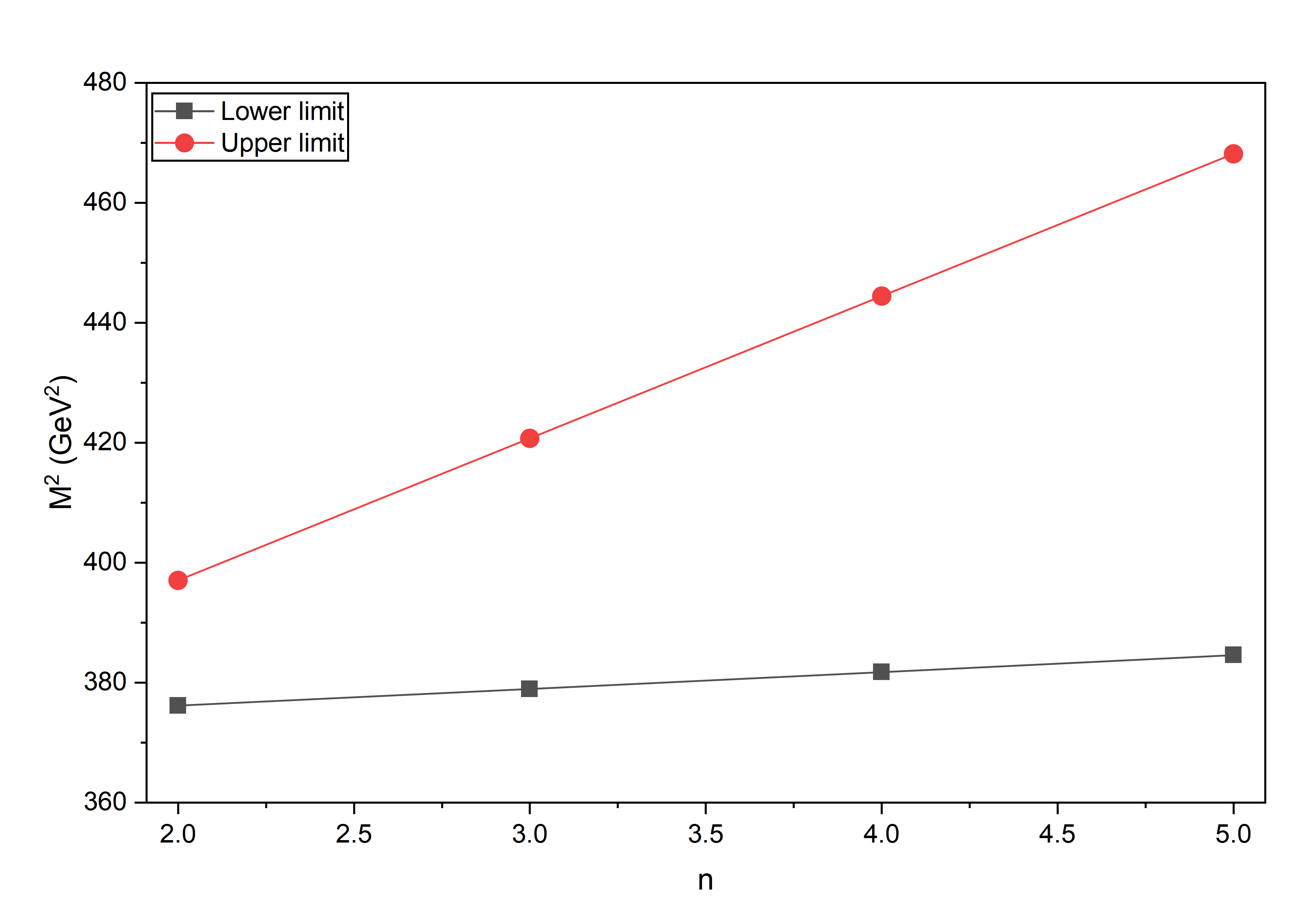}  
	\caption{Regge trajectory of $bb\bar{b}\bar{b}$ tetraqaurak for $S=0$ in $(n,M^2)$ plane.}
	\label{fig:regge trajectory for bbbb of S=0 in n plane}
\end{figure}

\section{Results and Discussion}

In this section, we analyze the mass spectra of the $ss\bar{b}\bar{b}$ and $bb\bar{b}\bar{b}$ tetraquark states obtained through the framework of Regge phenomenology. Utilizing a quasi-linear Regge trajectory approach, we derive both linear and quadratic mass relations that serve as consistency checks and provide bounds on the masses of tetraquark states.

To begin with, the ground state mass range of the $ss\bar{b}\bar{b}$ tetraquark system is estimated using mass inequalities derived using Regge phenomenology. The extracted mass intervals serve as a starting point for constructing Regge trajectories. The Regge slope parameters are then determined by fitting the $(J, M^2)$ trajectories for orbital excitations. These slopes, assumed to be nearly constant for a given flavor composition, enable us to predict the mass ranges of higher states for both $ss\bar{b}\bar{b}$ and $bb\bar{b}\bar{b}$ tetraquark configurations in $(J,M^2)$ plane. ( Table~\ref{tab:Slope} and Table~\ref{tab:SlopenM2} shows slopes of different tetraquarks in $(J,M^2)$ and $(n,M^2)$ planes respectively.)

In addition to orbital excitations, we extend our analysis to radial excitations by exploring the Regge trajectories in the $(n, M^2)$ plane. The slope parameters in this case are similarly extracted and used to generate the mass spectra of radial excitations. Our predictions for both ground and excited states are systematically compared with existing theoretical results from various models, including quark models and QCD sum rule approaches, where available.

The predicted mass spectra are consistent with the general expectations of Regge theory, and the mass ordering among different spin-parity configurations also aligns well with established phenomenological trends. In the following subsections, we present the detailed numerical results, Regge trajectory plots, and a comparative analysis with the literature.

\subsection{Doubly strange - doubly bottom ($ss\bar{b}\bar{b}$) and All-Bottom ($bb\bar{b}\bar{b}$) Mass Spectra in $(J,M^2)$ plane}

In this subsection, we analyze the predicted mass spectra of the $ss\bar{b}\bar{b}$ and $bb\bar{b}\bar{b}$ tetraquark systems in $(J,M^2)$ plane, evaluate their stability with respect to strong decays using the two-meson thresholds, and compare our results with existing theoretical studies. 

\subsubsection{Stability and Decay Channels}
Table~\ref{table:ssbb_tetraquarks} lists our calculated mass ranges for doubly strange- doubly bottom tetraquarks alongside two-meson thresholds.
The stability of tetraquark states can be inferred by comparing their masses with the corresponding two-meson decay thresholds. For the $ss\bar{b}\bar{b}$ system, the predicted ground state with $J^{PC}=0^{++}$ lies in the range $10.808$–$13.758$~GeV, with the lower end approximately $74$~MeV above the $B_s^0B_s^0$ threshold (10.734~GeV), suggesting that even the ground state is unstable against strong decay. Similarly, the $1^{+-}$ state is found in the range $10.826$–$13.766$~GeV, slightly above the $B_s^0B_s^{*}$ threshold (10.782~GeV), also indicating instability.

However, the proximity of the lower bound of these states to the respective thresholds suggests the possibility of narrow-width resonances or weakly bound states. All excited states of the $ss\bar{b}\bar{b}$ system clearly lie well above their corresponding thresholds and are expected to decay strongly into bottom-strange meson pairs.

Table~\ref{table:all_bottom_tetraquarks} presents the computed mass ranges of the all-bottom tetraquark states along with the corresponding two-meson threshold values.
For the fully-bottom $bb\bar{b}\bar{b}$ tetraquark, the lowest-lying states (e.g., $1^{--}$ in the range $19.366$–$19.687$~GeV) are compared with the $\Upsilon(1S)\chi_{b0}(1P)$ threshold (19.319~GeV), while others such as $2^{++}$ and $3^{--}$ also lie above respective bottomonium thresholds. These findings imply that the $bb\bar{b}\bar{b}$ tetraquarks are unstable against strong decays, though the predicted mass ranges are in some cases very close to the thresholds, indicating potential for narrow structures if binding energy is slightly underestimated.

\subsubsection{Comparison with Previous Studies}

Table~\ref{table:ssbb_tetraquarks_comparison} provides a comparison of our predicted $ss\bar{b}\bar{b}$ masses with previous theoretical results. For the $0^{++}$ ground state, our prediction (10.808–13.758~GeV) shows reasonable agreement with values from Refs.~\cite{Song2023,Braaten,Ebert2007,Wen2021,Luo2017,PhysRevD.102.034012}, which are generally clustered around 10.9–11.1~GeV. Similar consistency is observed for the $1^{+-}$ and $2^{++}$ states.

Table~\ref{table:all_bottom_tetraquarks_comparison} compares the $bb\bar{b}\bar{b}$ mass spectrum with various theoretical models. Our results align closely with those of Tiwari \textit{et al.}~\cite{Tiwari2021}, Faustov \textit{et al.}~\cite{sym14122504}, and Liu \textit{et al.}~\cite{PhysRevD.109.076017}. Slight deviations in higher orbital states may arise from differences in assumptions regarding diquark configurations, potential models, or parameter fitting. Notably, some studies such as Ref.~\cite{Chen:2024xyz} predict significantly lower masses for certain states, indicating model-dependence and uncertainties in the tetraquark sector.

\subsubsection{Spectral Systematics and Trends}

The predicted spectra exhibit systematic behavior across quantum numbers and excitation levels. As expected from Regge phenomenology, the mass increases with both spin and orbital anugular quantum numbers. For both systems, mass gaps between consecutive orbital excitations are found to be approximately linear in $J$, consistent with quasi-linear Regge trajectories. 

\subsubsection{Experimental Status}

Currently, no experimental confirmation exists for the $ss\bar{b}\bar{b}$ or $bb\bar{b}\bar{b}$ tetraquark states. Given the proximity of some predicted states to meson thresholds, particularly in the $bb\bar{b}\bar{b}$ sector, further experimental searches in the high-luminosity phase of LHC or future colliders with improved energy resolution could help confirm the existence of such exotic hadrons.

\subsection{Radial Excitations of $ss\bar{b}\bar{b}$ and $bb\bar{b}\bar{b}$ tetraquarks in the \texorpdfstring{$(n, M^2)$}{(n, M²)} Plane}

The mass spectra of the $ss\bar{b}\bar{b}$ and $bb\bar{b}\bar{b}$ tetraquark states are analyzed as functions of the radial quantum number $n$ in the $(n, M^2)$ plane. We present results for the $J^P = 0^{++}, 1^{+-},$ and $2^{++}$ channels up to $n=5$ in Tables~\ref{tab:ssbb_spectra_innm^2} and ~\ref{tab:bbbb_spectra_innm2}. The predicted masses are compared with existing theoretical models, where available.

For the $bb\bar{b}\bar{b}$ system, our results for the ground radial excitations ($n=2$ states) agree well with those reported in several previous studies. For instance, for the $0^{++}$ state, our predicted mass lies in the range $19.395$--$19.926$~GeV, which matches closely with the values given by Refs.~\cite{PhysRevD.109.076017,sym14122504,Bedolla2020,Lu2020,Zhao:2020,PhysRevD.104.014018,Mutuk2021,Ke2021}, most of which predict values around $19.3$--$19.8$~GeV. Similarly, for the $1^{+-}$ channel, our result of $19.402$--$19.927$~GeV is consistent with values in the $19.4$--$19.8$~GeV range reported by other models. In the $2^{++}$ channel, our prediction ($19.422$--$19.988$~GeV) also agrees with results from Refs.~\cite{PhysRevD.109.076017,sym14122504,Lu2020} and others, reinforcing the reliability of our method.

Our predicted masses for higher radial excitations ($n = 3$--$5$) follow a linear trend in the $(n, M^2)$ plane, consistent with Regge phenomenology. The comparison with available results for these excited states shows a good match where data exist. For example, the $3^1S_0$ and $3^3S_1$ states predicted around $19.47$--$20.51$~GeV show consistency with values from Refs.~\cite{sym14122504,Bedolla2020,Lu2020,Zhao:2020,PhysRevD.104.014018,Mutuk2021,Ke2021}. Some studies predict slightly higher or lower masses, but most remain within a margin of $\sim$200~MeV.

For the $ss\bar{b}\bar{b}$ system, no detailed comparative studies exist for excited states in the $(n, M^2)$ plane. However, our predicted radial spectra exhibit similar quasi-linear behavior as in the $bb\bar{b}\bar{b}$ case, with an overall downward shift in mass due to the presence of the lighter $s$ quarks. This trend is expected as the reduced mass of the system decreases, and is consistent with observations in other heavy-light tetraquark systems.

The slope of the radial Regge trajectories is slightly higher for the $ss\bar{b}\bar{b}$ system than for the $bb\bar{b}\bar{b}$ system, reflecting the inverse mass dependence of the Regge slope parameter. This is in accordance with the general expectation from Regge phenomenology and is similar to trends observed in quarkonium systems, where heavier systems tend to have smaller radial slopes~\cite{ref47,ref44,ref45,ref100,ref101,ref102,ref103,ref104,ref53}.

In summary, the mass spectra obtained in the $(n, M^2)$ plane are consistent with available results from various theoretical approaches, especially for the $bb\bar{b}\bar{b}$ system. Our predictions for $ss\bar{b}\bar{b}$ provide new insights into radial excitations and follow a systematic pattern in accordance with Regge behavior. The good agreement with previous studies and the consistent trend across spin and radial quantum numbers support the validity of our phenomenological framework.

\section{Conclusion}
The Regge phenomenological approach employed in this study provides a highly efficient yet robust framework for describing both orbital and radial excitations of fully heavy and heavy–strange tetraquark systems. Utilizing only a limited set of slope and intercept parameters, the model successfully reproduces the general pattern of mass hierarchies and level spacings, without relying on complex dynamical potentials or computationally intensive lattice QCD simulations. The linear nature of the Regge trajectories naturally reflects the near-uniform spacing observed in our calculated mass spectra and enables straightforward extrapolation to states with higher spin and radial quantum numbers.

Our predicted mass spectra demonstrate approximate linear behavior across increasing orbital and radial excitations, consistent with expectations from Regge theory. In the $(J, M^2)$ plane, we observed that many of the predicted states lie above their respective two-meson thresholds, suggesting that most of these tetraquark configurations may be unstable against strong decays. Nevertheless, some of the low-lying $ss\bar{b}\bar{b}$ and $bb\bar{b}\bar{b}$ states lie close to or below the relevant thresholds, indicating potential stability or narrow widths. These could be promising candidates for experimental observation in future searches.

Comparison with other theoretical approaches—including QCD sum rules, lattice QCD, and phenomenological quark models—shows reasonable agreement, especially for the lowest-lying S-wave states. Our results also exhibit similar patterns in radial excitations, and the extracted radial slope parameters align well with previously reported values in the literature. The overall consistency affirms the utility of Regge trajectories as an effective and predictive tool in exotic hadron spectroscopy.

Given the increasing experimental interest in fully-heavy and heavy-light tetraquarks, particularly at facilities like LHCb, CMS, and Belle II, we hope that our predictions will provide valuable guidance for future explorations of multiquark dynamics and the rich structure of QCD in the nonperturbative regime.

\section{Acknowledgment}
Vandan Patel acknowledges the financial assistance by University Grant Commision (UGC) under the CSIR-UGC Junior Research Fellow (JRF) scheme with Ref No. 231610186052.

\end{document}